\documentclass[prb,twocolumn]{revtex4-1}
\usepackage{amsfonts,amsmath,graphicx,natbib}
\pdfoutput=1
\newcommand{\BEQ}{\begin{equation}}
\newcommand{\BEA}{\begin{eqnarray}}
\newcommand{\EEQ}{\end{equation}}
\newcommand{\EEA}{\end{eqnarray}}

\newcommand\Tr{\mathop{\mathrm{Tr}}}
\newcommand\R\rangle
\renewcommand\L\langle
\renewcommand\r\right
\renewcommand\l\left

\newcommand\kv{\mathbf{k}}

\newcommand\om{\omega}
\newcommand\eps{\varepsilon}
\newcommand\s{\sigma}
\newcommand\up{\uparrow}
\newcommand\dn{\downarrow}
\newcommand\dg{\dagger}
\newcommand\Gm{\mathsf{G}}

\newcommand\tm{\mathsf{t}}
\newcommand\Sigmam{\mathsf{\Sigma}}
\newcommand\Gammam{\mathsf{\Gamma}}
\newcommand\kvt{{\tilde\kv}}
\newcommand{\Gc}{\mathcal{G}}

\begin{document}
\title{Bath optimization in the Cellular Dynamical Mean Field Theory}
\author{David S\'{e}n\'{e}chal}
\affiliation{D\'{e}partement de physique and Regroupement qu\'{e}b\'{e}cois sur les mat\'{e}riaux de pointe, Universit\'{e} de Sherbrooke, Sherbrooke, Qu\'{e}bec, Canada, J1K 2R1}
\date{\today}

\begin{abstract}
In the Cellular Dynamical Mean Field Theory (CDMFT), a strongly correlated system is represented by a small cluster of correlated sites, coupled to an adjustable bath of uncorrelated sites simulating the cluster's environment; the parameters governing the bath are set by a self-consistency condition involving the local Green function and the lattice electron dispersion.
Solving the cluster problem with an exact diagonalization method is only practical for small bath sizes (8 sites). In that case the self-consistency condition cannot be  exactly satisfied and is replaced by a minimization procedure.
There is some freedom in the definition of the `merit function' to optimize. We use Potthoff's Self-Energy Functional Approach on the one- and two-dimensional Hubbard models to gain insight into the best choice for this merit function. We argue that several merit functions should be used and preference given to the one that leads to the smallest gradient of the Potthoff self-energy functional. We propose a new merit function weighted with the self-energy that seems to fit the Mott transition in two dimensions better than other merit functions.
 
\end{abstract}


\maketitle

\section{Introduction}

The discovery of high-temperature superconductors in the late 1980's and the hypothesis that the mechanism of superconductivity in these materials is rooted in strong electron-electron interactions has stimulated theoretical investigation of lattice models of strongly-correlated electrons, such as the Hubbard model.
One of the early successes of this program was a new understanding of the metal-insulator transition in the Hubbard model using Dynamical Mean-Field Theory (DMFT)\cite{*[{For a review, see: }]Georges:1996}.
A central hypothesis behind DMFT, proven exact in the limit of infinite dimension,\cite{Metzner:1989fk} is that the momentum dependence of the electron self-energy $\Sigma(\mathbf{k},\om)$ may be neglected; DMFT focuses instead on the frequency dependence, which it determines approximately within a self-consistent procedure.
This is equivalent to replacing the original Hubbard model by an effective model in which a single site -- the ``impurity'' -- is embedded in the lattice through hybridization with a bath of uncorrelated orbitals.
The one-particle Green function $G(\om)$ for the correlated site then takes the form
\BEQ
G^{-1}(\om) = \om - \Gamma(\om) - \Sigma(\om)
\EEQ
where $\Sigma(\om)$ is the approximate electron self-energy and $\Gamma(\om)$ is the so-called {\em hybridization function} that incorporates the effect of the uncorrelated bath on the electron propagation.
The hybridization function is found by an interative procedure that involves (i) the solution of the impurity model and (ii) self-consistence between the electron Green function at the impurity site and the Green function $G(\mathbf{k},\om)$ constructed from the self-energy $\Sigma(\om)$ via Dyson's equation (more details on this below).

The importance of short-range antiferromagnetic fluctuations in Hubbard models and the possible existence of d-wave pairing has motivated the extension of DMFT to procedures where not only a site, but a finite cluster of sites, is embedded in the full lattice via a hybridization function.
This is further motivated by the strong momentum dependence of the self-energy inferred from photoemission experiments.\cite{Damascelli:2003}
The Dynamical Cluster Approximation (DCA)\cite{Hettler:1998,Hettler:2000} and the Cellular Dynamical Mean-Field Theory (CDMFT)\cite{Kotliar:2001} are two generalizations of DMFT to finite clusters, that take into account short-range spatial correlations.
Both have revealed antiferromagnetic order and d-wave pairing in the Hubbard model.\cite{Maier:2004fk,Maier:2005ec,Capone:2006bs,Haule:2007lh,Kancharla:2008,Civelli:2009fe}
They differ in that DCA is formulated in reciprocal space, by partitioning the Brillouin zone into a finite number of patches, whereas the CDMFT is formulated in direct space, by tiling the lattice into identical clusters with open boundary conditions.

\begin{figure}[b]
\centerline{\includegraphics[width=7.5cm]{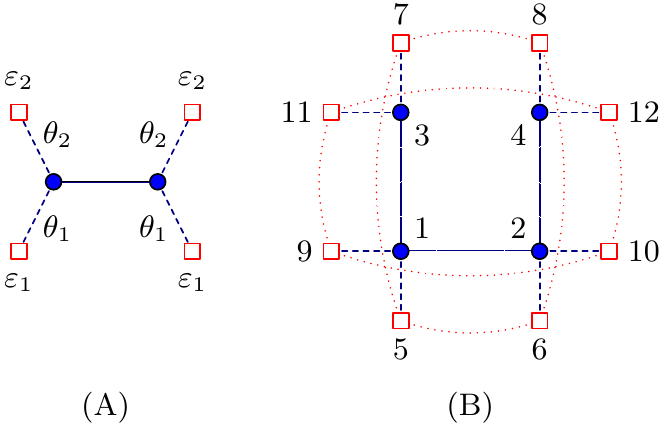}}
\caption{(Color online) Examples of clusters (blue circles) with finite baths (squares).
Cluster (A) is used to treat the one-dimensional Hubbard model and (B) the two-dimensional Hubbard model.}
\label{fig:bath}
\end{figure}

Solving the impurity problem in DMFT (or in its cluster extensions) may be done at finite temperature via Quantum Monte Carlo (QMC) or, as proposed in Ref.~\onlinecite{Caffarel:1994}, by the exact diagonalization (ED) of an associated Anderson Hamiltonian at zero-temperature.
The Monte Carlo approach has the advantage of simulating an effectively infinite bath and of providing temperature information.
On the other hand, it suffers from the infamous fermion sign problem, which makes convergence slow and the computational requirements important, even with the latest continuous-time algorithms free of discretization error.\cite{Gull:2008fu}
It is also very difficult to extract real-frequency dynamical information from it.
The pros and cons of the ED approach complement those of the QMC approach: (1) it is usually carried at zero temperature; (2) it provides real-frequency information; (3) it does not suffer from the sign problem but (4) it is limited to a small, discrete bath system.
This last characteristic is its most serious drawback, and the one we will deal with in this paper.

In Section II we review the CDMFT algorithm and explain how the restriction to a finite bath turns the self-consistency into an optimization problem whose solution depends on a choice of merit function. In Section III we review Potthoff's self-energy functional approach and assert that it provides the best possible hybridization function for the finite-bath problem. In Section IV we show the outcome of CDMFT calculations on the one- and two-dimensional Hubbard models, using various merit functions, and compare them with SFA results. We propose to select the merit function that minimizes the gradient of the Potthoff functional, in the cases where the SFA cannot be practically applied.

\section{The Cellular Dynamical Mean Field Theory}

In CDMFT, the full lattice Hubbard Hamiltonian $H$ is replaced by the following cluster Hamiltonian:
\BEQ
\begin{split}
\label{eq:cdmft1}
H' =&-\sum_{i,j,\s}t_{ij}c_{i\s}^\dg c_{j\s} + U\sum_{i}n_{i\up}n_{i\dn}\\
&+\sum_{i,\mu,\s}\theta_{i\mu}(c_{i\s}^\dg a_{\mu\s} + \mathrm{H.c.}) + \sum_{\mu,\s} \eps_\mu a_{\mu\s}^\dg a_{\mu\s}
\end{split}
\EEQ
where $c_{i\s}$ annihilates an electron of spin $\s$ on a physical site labelled $i$, and $a_{\mu\s}$ annihilates an electron of spin $\s$ on a bath orbital labelled $\mu$.
The bath is parametrized by the energy of each orbital ($\eps_\mu$) and by the bath-cluster hybridization matrix
$\theta_{i\mu}$ (we assume spin-independence for simplicity; this would not be the case in a treatment of antiferromagnetism).
Note that `bath site' is a misnomer, as bath orbitals have no physical position assigned to them.
The clusters and baths used in this work are shown on Fig.~\ref{fig:bath}.
Our convention is to add to the above Hamiltonian a chemical potential term $-\mu\hat N$, where $\hat N$ is the total number of electrons in the cluster and the bath.

Note that a more general parametrization of the systems illustrated on Fig.~\ref{fig:bath} is possible\cite{Koch:2008ix,Liebsch:2009bv}; in particular, the most general bath made of 8 orbitals and compatible with the discrete symmetry of the cluster of Fig.~\ref{fig:bath}B would require 16 parameters.
But our purpose is to illustrate the relationship between the CDMFT-ED procedure and Potthoff's self-energy functional approach, and accordingly we choose to limit the size of the parameter set.
In practice, cluster symmetries are used in the exact diagonalization in order to accelerate convergence and cut memory costs.\cite{Koch:2008ix,Senechal:2008gh}

The effect of the bath on the electron Green function is encapsulated in the so-called hybridization function
\BEQ\label{eq:hybridization}
\Gamma_{ij}(\om) = \sum_\mu \frac{\theta_{i\mu}\theta^*_{j\mu}}{\om-\eps_\mu}
\EEQ
which enters the cluster electron Green function as
\BEQ\label{eq:hybridization2}
\Gm'{}^{-1} = \om - \tm' - \Gammam(\om) - \Sigmam(\om)
\EEQ
where we hide site and spin indices behind a matrix notation for the cluster one-body terms ($\tm'$, including chemical potential), the hybridization function ($\Gammam$), the self-energy ($\Sigmam$) and the cluster Green function ($\Gm'$).

\begin{figure}[cbt]
\centerline{\includegraphics[width=6cm]{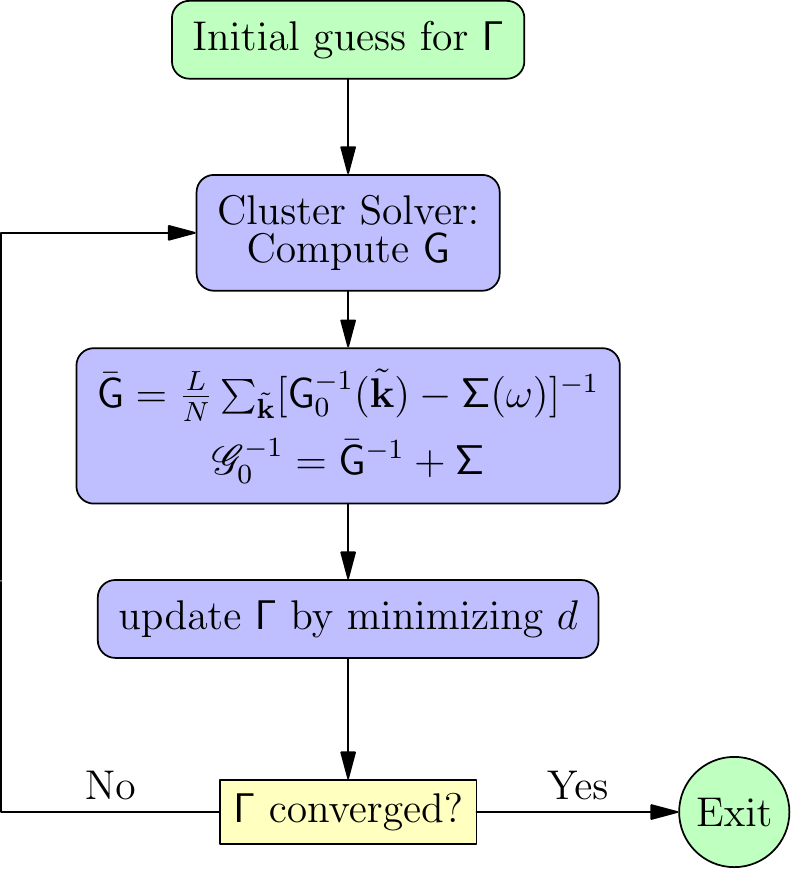}}
\caption{(Color online) The CDMFT algorithm with an exact diagonalization solver.}
\label{fig:algo}
\end{figure}

The basic computational task of DMFT approaches is to find the best possible embedding of the cluster into the original lattice; that is, to find the best possible value of the bath parameters.
In CDMFT, this is accomplished via a self-consistency condition, using the following iterative algorithm, 
summarized in Fig.~\ref{fig:algo}:
\begin{enumerate}
\item Start with a guess value of the bath parameters $(\theta_{i\mu},\eps_\mu)$, that define the hybridization function (\ref{eq:hybridization}).
\item Calculate the cluster Green function $\Gm'(\om)$ with the Exact diagonalization solver and extract the cluster self-energy $\Sigmam(\om)$.
\item Construct the momentum-dependent Green function $\Gm(\kvt,\om)$ from that self-energy and project it back on the cluster via a Fourier transform:
\BEQ\label{eq:Gaverage}
\bar\Gm(\om)=  \int_{\kvt}\frac{1}{\Gm_{0}^{-1}(\kvt)-\Sigmam(\om)}
\EEQ
where $\int_\kvt$ stands for an integral over the reduced Brillouin zone, along with the appropriate factors of $2\pi$.
\item Calculate the matrix
\BEQ
\Gc_{0}^{-1}(\om)=\bar\Gm^{-1}+\Sigmam(\om)
\EEQ
\item Choose new bath parameters that make the combination $\om -\tm'-\Gammam(\om)$ as close as possible to $\Gc_{0}^{-1}$.
Since we have a finite number of bath parameters at our disposal, this self-consistency condition cannot be fulfilled for all frequencies, but only optimized.
This is done by minimizing the distance function
\BEQ\label{eq:dist}
d=\sum_{\om,i,j}W(\om)\l\vert\l(\om -\tm'-\Gammam(\om)-\Gc_0^{-1}(\om)\r)_{ij}\r\vert^{2}
\EEQ
over the set of bath parameters. 
Changing the bath parameters at this step does not require a new solution of the Hamiltonian $H'$, but merely a recalculation of the hybridization matrix (\ref{eq:hybridization}).
\item Go back to step (2) with the new bath parameters obtained from this minimization, and repeat until they are converged.
\end{enumerate}

In practice, the distance function (\ref{eq:dist}) can take various forms, depending on the heuristic weight function $W(\om)$.
So far, ad hoc criteria and intuition have been used to choose the weight function.
Those that are benchmarked in this paper are listed in Eq.~(\ref{eq:W}) below.
The sum over frequencies in Eq.~(\ref{eq:dist}) is carried over a set of equally spaced Matsubara frequencies, defined by some fictitious inverse temperature $\beta$, typically ranging from 20 to 200 (in units of $t^{-1}$).

\section{The SFA approach}

There is a more fundamental way to find the best possible value of the bath parameters: Potthoff's self-energy functional approach (SFA).\cite{Potthoff:2003b}
Let us summarize this approach here:
The self-energy $\Sigmam$ of any system obeys a variational principle in the space of all possible self-energies:
\BEQ\label{eq:varia1}
\frac{\delta\Omega_{\tm,U}[\Sigmam]}{\delta\Sigmam}=0
\EEQ
where the functional $\Omega_{\tm,U}[\Sigmam]$ is given by
\BEQ\label{eq:sef1}
\Omega_{\tm,U}[\Sigmam]=F_U[\Sigmam]-\Tr\ln(-\Gm_{0\tm}^{-1}+\Sigmam)
\EEQ
$\Gm_{0\tm}$ is the noninteracting Green function and $F_U[\Sigmam]$ is the Legendre transform of the Luttinger-Ward functional $\Phi[\Gm]$:
\BEQ
F_U[\Sigmam]=\Phi_U[\Gm]-\Tr(\Sigmam\Gm)\qquad \Sigmam\equiv\frac{\delta\Phi_U}{\delta\Gm}
\EEQ
The value of the functional $\Omega_{\tm,U}[\Sigmam]$ evaluated at the physical self-energy is precisely the grand potential $\Omega$ of the system.
Moreover, the functional $F_U$ is universal, in the sense that its form does not depend on the one-body part $\tm$ of the Hamiltonian, but only on the interaction part.
This means that it can be evaluated by considering a different Hamiltonian, $H'$, that shares the same interaction part with the original Hamiltonian $H$, but has a different one-body part.
In particular, the cluster Hamiltonian (\ref{eq:cdmft1}) falls in that category.
This allows us to extract the value of the functional $F$ evaluated at the physical self-energy of $H'$ and to calculate
\BEQ\label{eq:sef2}
\Omega_\tm[\Sigmam] = \Omega'+\Tr\ln(-\Gm')-\Tr\ln(-\Gm)
\EEQ
where $\Gm$ stands for $(\Gm_{0\tm}^{-1}-\Sigmam)^{-1}$ and $\Omega'$ is the grand potential associated with the cluster Hamiltonian $H'$.
A more explicit expression is
\BEQ\label{eq:sef3}
\Omega_\tm[\Sigmam] = \Omega' - T\sum_\om\int_\kvt \ln\det\l[1-(\tm_\kvt-\tm')\Gm'(\kvt,\om)\r]
\EEQ
where $\tm_\kvt$ is the one-body matrix of the original system (here expressed as a matrix over cluster site indices and a function over the reduced Brillouin zone) and $\tm'$ is the one-body matrix of the cluster Hamiltonian $H'$.
$T$ is the absolute temperature and the sum is carried over Matsubara frequencies (at zero temperature this translates into an integral over the imaginary-frequency axis).

The bath system is of course assumed to be decoupled from the cluster in the original Hamiltonian $H$.
With a finite bath, the functional $\Omega_\tm$ becomes effectively an ordinary function of the bath parameters $\theta_{i\mu}$ and $\eps_\mu$.
An objective answer to the question of what are the optimal values of the bath parameters is obtained by solving the stationary condition $\partial\Omega_\tm/\partial h=0$, where $h$ stands for any one of the bath parameters.
This condition may be further explicited as
\BEQ\label{eq:EulerVCA}
\sum_\om\Tr\l\{  \Big[\Gm'^{-1}(\om) -\bar\Gm^{-1}(\om) \Big]\cdot\frac{\partial\Sigmam'(\om)}{\partial h} \r\} =0.
\EEQ
where $\bar\Gm(\om)$ stands for the Brillouin zone averaged Green function defined in Eq.~(\ref{eq:Gaverage}).
Note that the hybridization function $\Gammam(\om)$ enters $\Gm_{0\tm'}^{-1}$, as can be seen from Eq.~(\ref{eq:hybridization2}).
The distance function (\ref{eq:dist}), by contrast, may be recast as
\BEQ\label{eq:dist2}
d = \sum_\om W(\om) \Tr  \Big[\Gm'^{-1}(\om) -\bar\Gm^{-1}(\om) \Big]^2
\EEQ
The condition (\ref{eq:EulerVCA}) does not entail $d=0$, since the latter can only be satisfied with an infinite number of bath parameters, $d$ being a sum of positive-definite contributions.
The SFA solution is therefore not self-consistent ($\Gm'(\om)\not\equiv\bar\Gm(\om)$), but it is, in a variational sense,\cite{Potthoff:2005} the best possible approximation to the original Hamiltonian by a cluster-bath system.

\begin{figure*}
\centerline{\includegraphics[width=16cm]{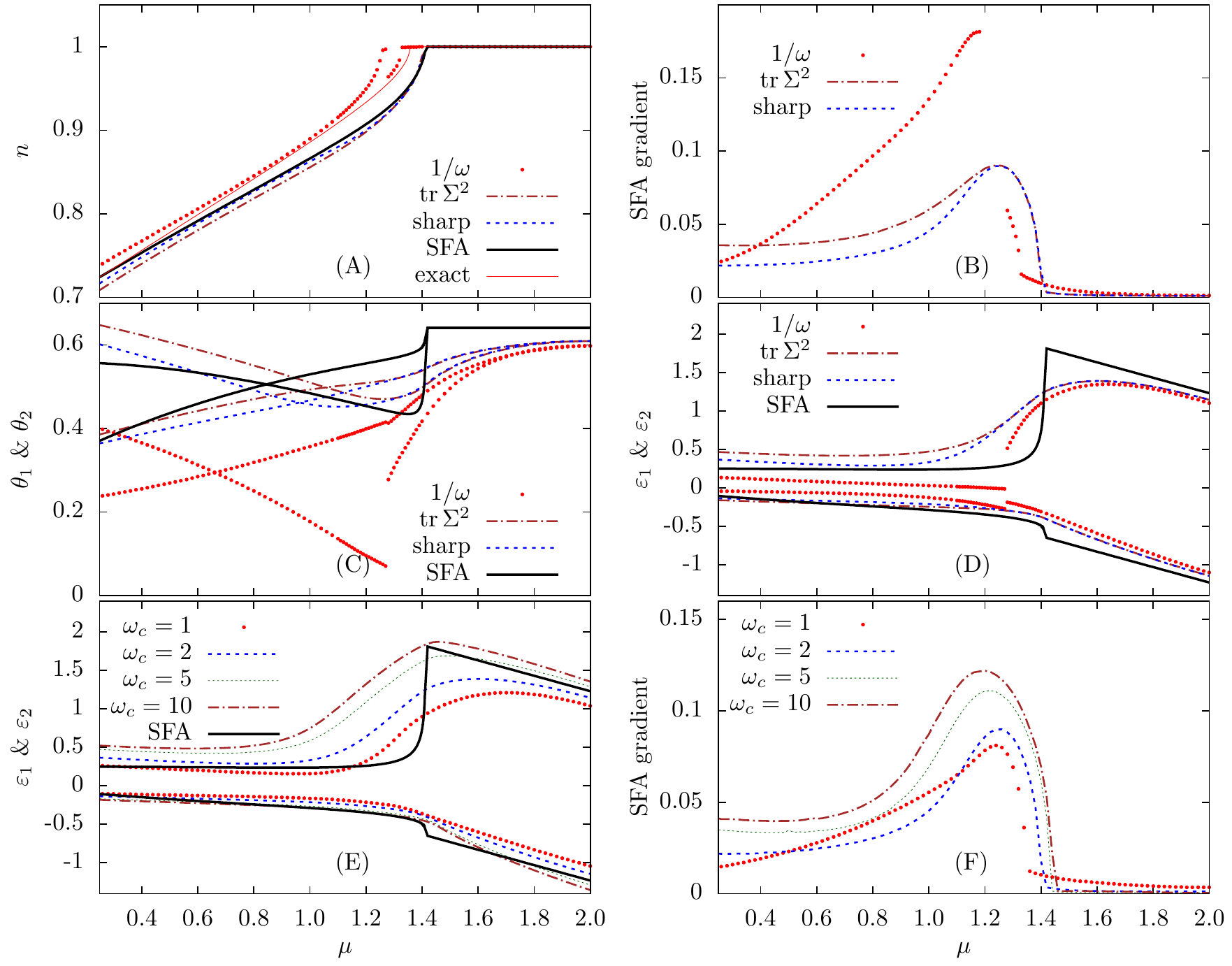}}
\caption{(Color online) Plots illustrating the CDMFT and SFA solutions from the cluster of Fig.~\ref{fig:bath}A, for the one-dimensional Hubbard model, all expressed as a function of the chemical potential $\mu$.
Panel (A): Electron density $n$ (the exact Lieb \& Wu result is also shown). 
Panel (B): SFA gradient associated with the various CDMFT solutions.
Panel (C) and (D): Hybridization parameters $\theta_{1,2}$ and bath energies $\eps_{1,2}$ for the weight functions (\ref{eq:W}) and the SFA solution.
Panel (E) and (F): Bath energies $\eps_{1,2}$ for a sharp cutoff (\ref{eq:W}a) with various values of $\om_c$, and the corresponding values of the SFA gradient. Unless indicated otherwise, the cutoff was set at $\om_c=2$ and the fictitious inverse temperature at $\beta=100$.}
\label{fig:1D}
\end{figure*}

If we think of the expression between brackets in (\ref{eq:dist2}) as forming a vector in both frequency and site index space, then the CDMFT tries to make that vector as small as possible, whereas the SFA tries to make it orthogonal to the vector $\partial\Sigmam'/\partial h$, {\em or} to make it vanish, the latter being impossible with a finite bath.
Whereas a perfect solution ($d=0$) of the self-consistency condition would automatically satisfy Condition (\ref{eq:EulerVCA}), the converse is not necessarily true.
The SFA is not the only functional formulation from which DMFT can be derived (see Ref.~\onlinecite{Kotliar:2006kx} for an extensive discussion). However, it has the distinction of being based on an exact evaluation of the functional, albeit in a restricted space of parameters.\cite{Potthoff:2005}

Detailed benchmarks of the SFA approach against the exact solution of half-filled the one-dimensional Hubbard model have been performed in Ref.~\onlinecite{Balzer:2008}, for a large variety of clusters (with and without baths).
In this work we focus instead on comparing the SFA approach with CDMFT, at, and away from, half-filling.

\section{Results}

Fig.~\ref{fig:bath}A illustrates a simple bath-cluster system that can be used to approximate the one-dimensional Hubbard model: two cluster sites and four bath sites, with two hybridization parameters $\theta_{1,2}$ and two bath energies $\eps_{1,2}$. 
The presence of four bath sites (as opposed to two) is required in order to have particle-hole symmetry at half-filling.
This bath system was used, for instance, in Refs~\onlinecite{Capone:2004qo,Go:2009fu}.

From a computational point of view, solving the SFA condition (\ref{eq:EulerVCA}) is more difficult to carry out than the self-consistent CDMFT algorithm of Fig.~\ref{fig:algo}.
More instances of the impurity solver must be called, and the optimization of the functional (\ref{eq:sef2}) with respect to the bath parameters must be very carefully done: it requires great numerical precision and is prone to instabilities, because of the relatively weak dependence of $\Omega_\tm$ on the bath parameters.
Thus, solving the SFA condition (\ref{eq:EulerVCA}) can only carried out in practice on small systems with few variational parameters, but it sheds light on the proper choice of weighting function $W$ to be used in the CDMFT self-consistent procedure.

We have tested the following weight functions $W(\om)$ against the SFA results:
\begin{subequations}\label{eq:W}\begin{align}
W(i\om)&=1 \text{ within } \om\in[0,\om_c] \text{ (sharp cutoff)} \\
W(i\om) &=1/\om \text{ (extra weight to low frequencies)}\\
W(i\om) &= \Tr|\Sigmam^2(i\om)| \text{ within } \om\in[0,\om_c]
\end{align}\end{subequations}
All of these functions have a finite support between $\om=0$ and some cutoff frequency $i\om_c$, and are evaluated on a grid of Matsubara frequencies defined by a `fictitious' temperature $1/\beta$ (recall that the ED solver is used strictly at zero-temperature in this work).
Using a range of frequencies along a segment parallel to (and slightly above) the real axis has also been tried, but gives very unreliable results, presumably because the landscape of the distance function (\ref{eq:dist}) is much more complicated, as the zeros (and poles) of the Green function are located on the real axis.
Other weight functions $W$ have also been benchmarked, for instance by putting greater emphasis on low frequencies ($W(\om)=1/\om^2$), or less emphasis ($W(\om)=1/\sqrt{\om}$), or proportional to $|\Tr\Sigmam|$ instead of $\Tr|\Sigmam|^2$. They bring nothing qualitatively different than the choices (\ref{eq:W}) that we benchmark here.

Note that these various distance functions are all independent of the choice of basis used for bath or site orbitals:
The factor multiplying $W$ in (\ref{eq:dist}) is a matrix trace, and so is choice (\ref{eq:W}c).
Choice (\ref{eq:W}b) is motivated by the desire to give much more weight to low frequencies, and 
Choice (\ref{eq:W}c) by the desire to give more weight to frequencies with a large self-energy.
The cutoff $\om_c$ may also be chosen so as to weigh more low frequencies.

Fig.~\ref{fig:1D} summarizes the benchmarks we have conducted on the one-dimensional Hubbard model, using the cluster illustrated on Fig.~\ref{fig:bath}A, with $U=4$ and nearest-neighbor hopping $t=1$.
Panels (C) and (D) show the optimal value of the bath parameter $\theta_{1,2}$ and $\eps_{1,2}$ as a function of chemical potential (within each pair, the two parameters are of course interchangeable, and so there is no point in labelling them separately).
The SFA result is non-analytic at a value of the chemical potential ($\mu_c\approx1.4$) corresponding to the edge of the gap in the one-dimensional Hubbard model. In the range $\mu\in[\mu_c,U/2]$, the SFA values of $\eps_1+\mu$ and $\theta_1$ are constant, as they should be since the physical state of the system is the same for all values of $\mu$ within the Mott gap (the same applies to $\eps_2$ and $\theta_2$).
This supports our view that the SFA provides the best possible values of the bath parameters.
On the other hand, the bath parameters obtained from CDMFT with a sharp cutoff (\ref{eq:W}a) or a self-energy weight (\ref{eq:W}c) are analytic at $\mu_c$, even though they follow the general trend of the SFA solution. 
The weight function $\sim1/\om$ (\ref{eq:W}b) leads to solutions that show some non-analycity, but depart more from the SFA solution than the other two.

\begin{figure}[ctb]
\centerline{\includegraphics[width=8cm]{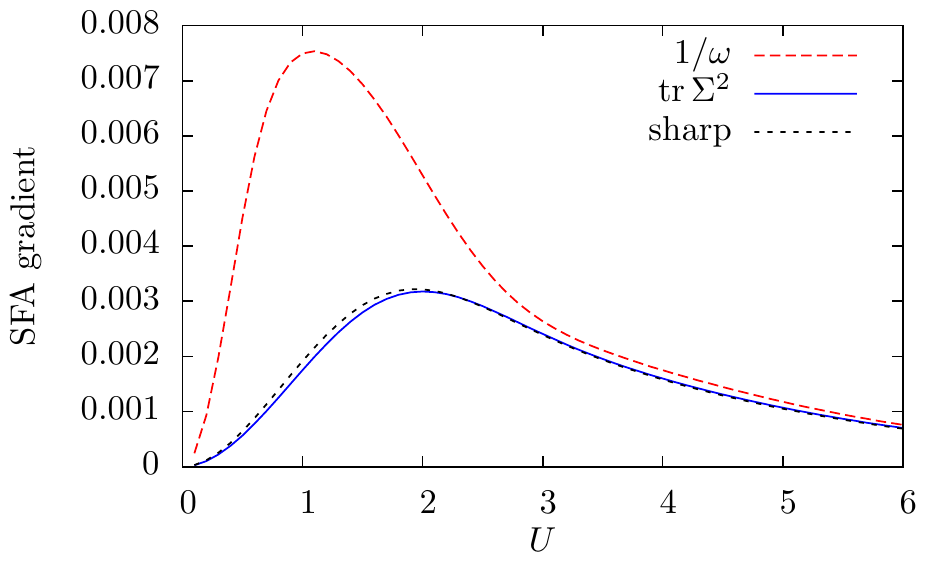}}
\caption{(Color online) Gradient of the SFA functional for the various CDMFT weight functions (\ref{eq:W}). Same system as Fig.~\ref{fig:1D}, but this time at half-filling, as a function of $U$.}
\label{fig:hf}
\end{figure}
\begin{figure}[ctb]
\centerline{\includegraphics[width=8cm]{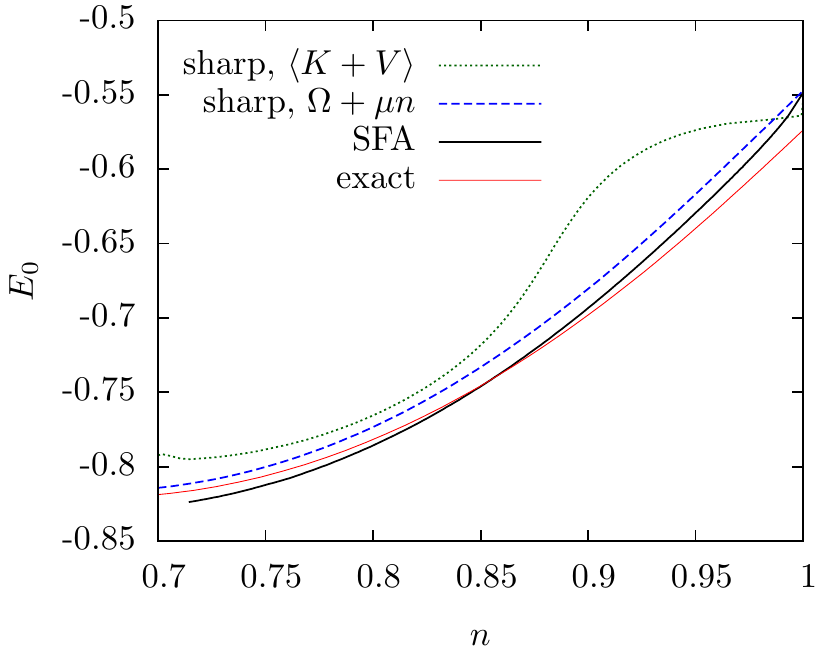}}
\caption{(Color online) Ground state energy density of the one-dimensional Hubbard model, estimated in various ways (see text). The exact result is shown for comparison.}
\label{fig:ener}
\end{figure}

Fig.~\ref{fig:1D}B shows the norm of the gradient of the SFA functional (\ref{eq:sef3}) evaluated at the CDMFT solutions found using the weight functions (\ref{eq:W}) (this gradient is precisely zero at the SFA solution). This can be used as a measure of the departure from the SFA solution, even in cases where the SFA solution is not known.
Again, the weight functions (\ref{eq:W}a) and (\ref{eq:W}c) appear to be the most sensible, while the one that enhance low frequencies (\ref{eq:W}b)  has most of the time the largest gradient.

Fig.~\ref{fig:1D}E shows the CDMFT values of the bath energies $\eps_{1,2}$, using a sharp cutoff (\ref{eq:W}a), for different values of the cutoff frequency $\om_c$. Panel F shows the SFA gradient for the same set of data.
This clearly shows that $\om_c$ should be small enough, but that $\om_c=1$ is too small and gives non optimal results in the gapped region.
As a rule, the value $\om_c=2$ provides the best results.
One could also display the same type of analysis as a function of fictitious temperature $\beta^{-1}$.
In that case, one can show that the value $\beta=100$ (i.e., a fictitious temperature at 1\% of the hopping amplitude $t$) is a good choice, in terms of smoothness and ease of calculation; this is the value that was used in all other plots of this paper. On the other hand, $\beta=20$ is definitely too low.

Finally, Fig.~\ref{fig:1D}A shows the CDMFT values of the electron density $n$, using the weight functions (\ref{eq:W}), as well as the value obtained from the SFA solution and, this time, from Lieb and Wu's exact solution of the one-dimensional Hubbard model\cite{Lieb:1968fk}.
Again, the weight functions (\ref{eq:W}a) and (\ref{eq:W}c) are closest on average to the SFA solution. Note however that the latter does not coincide with the exact solution and that the weight function (\ref{eq:W}b) yields a solution that is sometimes closer to the exact solution.
The SFA solution would move closer to the exact solution if either the number of sites or the bath size were increased.
But we argue that it provides the best solution for the cluster and bath used here, and that it should be the standard against which the different CDMFT solutions are compared. We view the occasional close proximity of a CDMFT solution to the exact solution as accidental.
In particular, the solutions obtained from the low-frequency weight function (\ref{eq:W}b) can be tuned to yield the correct value of the critical chemical potential $\mu_c$ by adjusting $\beta$, but that does not mean a convergence towards the exact value as $\beta^{-1}\to0$.

Fig.~\ref{fig:hf} shows the same type of comparison, this time at half-filling, as a function of $U$. Again, the sharp cutoff (\ref{eq:W}a) and the self-energy weight (\ref{eq:W}c) stand out as the best choices. Note that the gradient goes to zero in the $U/t\to 0$ and $t/U\to0$ limits, which is natural given that quantum cluster methods such as CDMFT become exact in these limits.

Fig.~\ref{fig:ener} shows estimates of the ground state energy density $E_0$ of the 1D Hubbard model as a function of density $n$. The exact Lieb \& Wu result is shown for comparison, as well as the SFA value obtained from the optimal value $\Omega$ of the functional (\ref{eq:sef3}) by the relation $E_0 = \Omega + \mu n$. For the sharp cutoff (\ref{eq:W}a), we provide estimates of the ground state energy density obtained in two ways: (1) by calculating the average $\L K+V \R$, where the average $\L K\R$ of the kinetic energy is calculated from the lattice Green function $G(\om,\kv)$, and the average $\L V\R$ of the potential energy is calculated from the ground state double occupancy at a cluster site; (2) by calculating the functional (\ref{eq:sef3}) and adding $\mu n$.
We conclude from this graph that method (2) provides a better estimate of the ground state energy than method (1), even though the CDMFT solutions are not exact solutions of the variational equations (\ref{eq:EulerVCA}).
The same conclusion is reached with the other weight functions (\ref{eq:W}).

\begin{figure}[ctb]
\centerline{\includegraphics[scale=0.9]{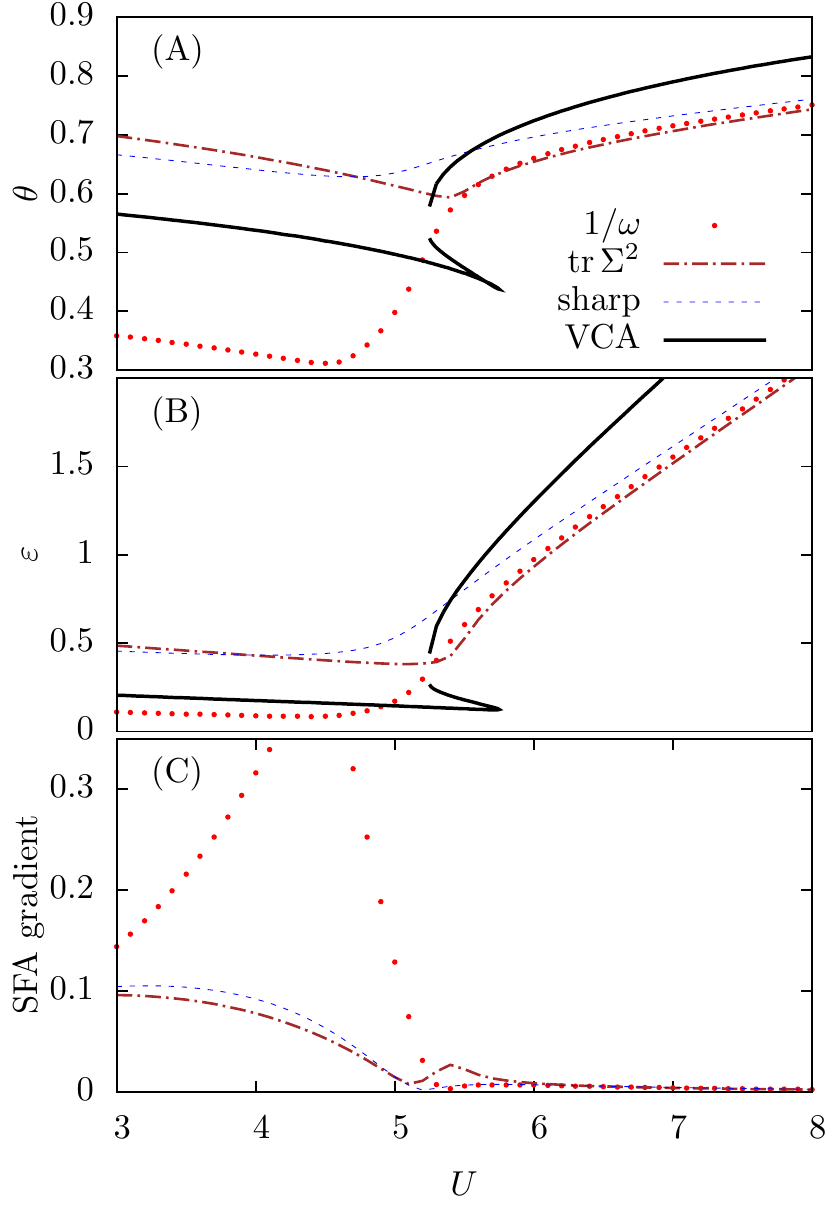}}
\caption{(Color online) Top: hybridization parameter $\theta$ for the half-filled, two-dimensional Hubbard model, with weight functions (\ref{eq:W}a), compared with the SFA result taken from Ref.~\onlinecite{Balzer:2009kl}. Middle panel: same for the bath energy $\eps$. Bottom panel: the SFA gradient for the same solutions.}
\label{fig:2D-hf}
\end{figure}
\begin{figure}[ctb]
\centerline{\includegraphics[scale=0.9]{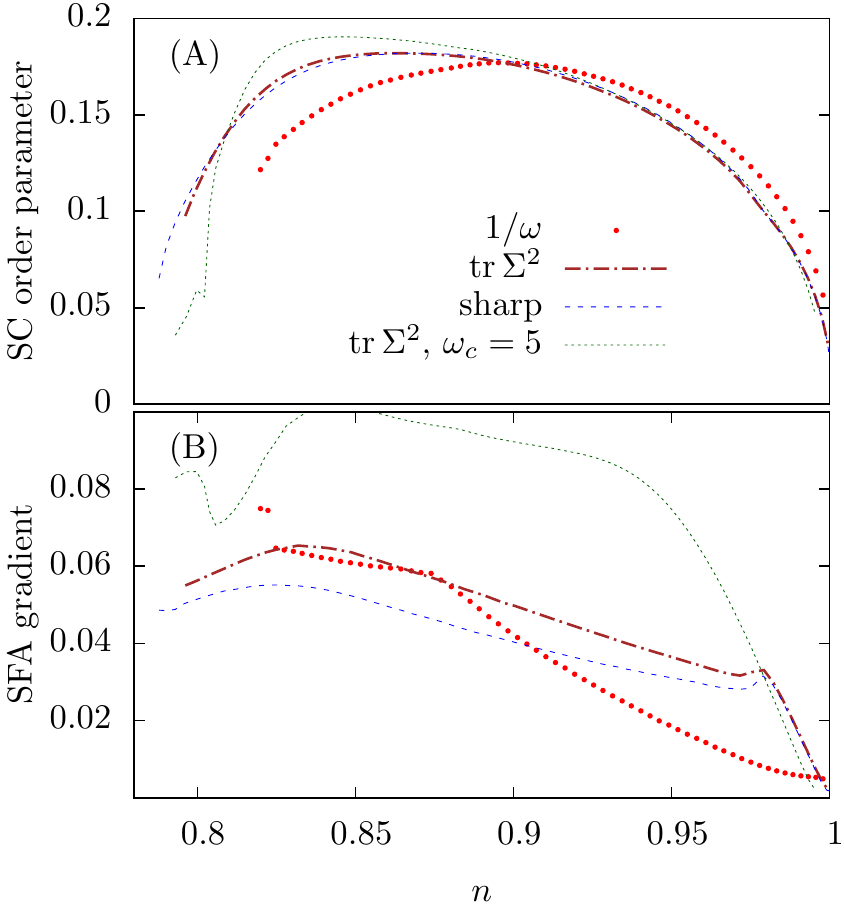}}
\caption{(Color online) Top: d-wave order parameter as a function of density in the two-dimensional Hubbard model, obtained from CDMFT, with the weight functions (\ref{eq:W}). Bottom: the SFA gradient for the same solutions.}
\label{fig:2D}
\end{figure}

Let us turn to the two-dimensional Hubbard model, again with nearest-neighbor hopping only.
The cluster-bath system is illustrated on Fig.~\ref{fig:bath}B.
We will start with a discussion of the half-filled system, in which case only two bath parameters are necessary because of particle-hole symmetry: a hybridization $\theta$ (dashed lines on Fig.~\ref{fig:bath}B), and a bath energy $\pm\eps$ (positive on bath sites labelled 5 through 8, negative on the others).
The SFA solution for this system was obtained in Ref.~\onlinecite{Balzer:2009kl} as a function of $U$ and revealed a Mott transition with two critical values of the Coulomb repulsion ($U_{c1}$ and $U_{c2}$).
This is illustrated by the full black curve on Fig.~\ref{fig:2D-hf}: there is an insulating solution at strong $U$ that overlaps with a metallic solution as smaller $U$, and an `unstable' solution linking the two, as one would expect for a first-order transition described, e.g., by Landau's theory of phase transitions.
The CDMFT solutions that are found for the same problem are shown on Fig.~\ref{fig:2D-hf} for three possible weight functions.
The Mott transition is visible through an upturn of the bath parameters, but no hysteresis was observed: sweeping $U$ upwards or downwards did not make any difference. The bottom panel of Fig.~\ref{fig:2D-hf} shows the SFA gradient calculated from the three CDMFT solutions.
Overall, the weight function (\ref{eq:W}c) has the lowest gradient, except exactly at the transition; paradoxically, this weight function also best describes the transition: the minimum of $\theta$ and the sharp upturn in $\eps$ occur right between the $U_{c1}$ and $U_{c2}$ found in the SFA, and the changes observed near the transistions are the sharpest of the three weight functions used.
At the Mott transition, one would naturally expect a hysteresis loop from CDMFT solutions, or a local increase in the SFA gradient due to shift of the solution from the vicinity of the metallic SFA solution to the vicinity of the insulating SFA solution.
The solution found using the self-energy weight (\ref{eq:W}c) does precisely that.

Note that the $1/\om$ weight (\ref{eq:W}b) seems particularly inadequate for the metallic solution, which runs against the intuition that the $1/\om$ weight would better describe a state with low-energy states like a metal. This is also true of the one-dimensional system described in Fig.~\ref{fig:1D}.
However, this is naturally understood in the context of the variational equations (\ref{eq:EulerVCA}): in the low-frequency limit, the self-energy of a Fermi liquid vanishes whereas that of a Mott insulator or pseudo-gapped system is large. Thus, if the appropriate weight is to be based somehow on the self-energy, the $1/\om$ weight function should be more appropriate for a Mott insulator, not a metal.

Finally, we probe d-wave superconductivity in the two-dimensional Hubbard model.
This is done like in Refs \onlinecite{Capone:2006bs,Kancharla:2008}, using the cluster-bath system illustrated on Fig.~\ref{fig:bath}B.
In this case we used six bath parameters: a pair $(\eps_1,\theta_1)$ of bath energy and hybridization for the `first' bath, made up of the orbitals labelled 5 through 8; a similar pair $(\eps_2,\theta_2)$ for the `second' bath, made up of the orbitals labelled 9 through 12;
two pairing parameters $(d_1,d_2)$, multiplying pairing operators $\hat d_{1,2}$ with d-wave symmetry, symbolically represented by the dashed curves on Fig.~\ref{fig:bath}B. 
The exact expression of $\hat d_1$ is
\BEQ\begin{split}
\hat d_1 &= a_{5\up}a_{6\dn} + a_{6\up}a_{5\dn} + a_{7\up}a_{8\dn} + a_{8\up}a_{7\dn} \\
&\quad - a_{5\up}a_{7\dn} - a_{7\up}a_{5\dn} - a_{6\up}a_{8\dn} - a_{8\up}a_{6\dn} + \text{ H.c}
\end{split}\EEQ
and $\hat d_2$ has the corresponding expression for bath sites 9 to 12.
The introduction of the pairing parameter breaks the conservation of particle number in the cluster-bath system, and anomalous averages may be nonzero, which is taken as the signature of superconductivity in the system.
in particular, what we call the d-wave order parameter is the ground-state average of the operator
\BEQ
\hat D = \int_\kv (\cos(k_x)-\cos(k_y))\Big(c_{\kv\up}c_{-\kv\dn} + c_{-\kv\up}c_{\kv\dn} + \text{ H.c}\Big)
\EEQ
where $\int_\kv$ stands for an integral over the original Brillouin zone (along with the appropriate factors of $2\pi$).
The average $\L\hat D\R$ can be calculated from the lattice Green function $G(\om,\kv)$ obtained from the CDMFT solution.

Fig.~\ref{fig:2D}A shows the d-wave order parameter for the two-dimensional Hubbard model with nearest-neighbor hopping $t=1$ and on-site repulsion $U=8$, as a function of electron density. We only show the hole-doped side, since this system is particle-hole symmetric.
The outcome of four CDMFT weight functions is plotted. 
The three weight functions (\ref{eq:W}) are used with $\om_c=2$, and in addition a higher cutoff ($\om_c=5$) is used with weight function (\ref{eq:W}c).
At this point this system is beyond the reach of the SFA, as it presents considerable numerical challenges.
However, this does not prevent us from computing the SFA gradient once the CDMFT solution is found, and this can be used as a heuristic measure of the proximity to the unknown SFA solution.
All CDMFT solutions found have superconductivity, and roughly in the same range, so the choice of weight function is largely a quantitative, not qualitative, issue.
The use of a larger cutoff ($\om_c=5$) can be rejected on the basis that it displays the largest SFA gradient of the set (this agrees again with the conclusions drawn from the one-dimensional system).
The $1/\om$ weight function (\ref{eq:W}b) seems more appropriate in the underdoped region: it has the lowest SFA gradient there and the largest SC order parameter;
this is also where the spectral gap is largest (not shown).
On the other hand, the sharp cutoff weight (\ref{eq:W}a) is more adequate in the overdoped region, where the SC gap is smaller.
This again confirms that the $1/\om$ weight should not be used in a metallic of low-gap phase.\\

\section{CONCLUSION}

In the Cellular Dynamical Mean Field Theory, the self-consistency condition $\Gm(\om)\equiv\bar\Gm(\om)$ cannot be exactly satisfied when using an exact diagonalization solver, because of the small size of the bath.
In other words, the distance function (\ref{eq:dist}) cannot be made to vanish, but can only be minimized.
In that case, an ambiguity arises because of the arbitrariness in the choice of the weight function $W(\om)$, and this ambiguity translates into a variety of solutions with sometimes important quantitative differences.
We argued that Potthoff's Self-Energy Approach provides the best possible solution for the CDMFT bath parameters; in particular, it is non-analytic at the critical value $\mu_c$ of the chemical potential that separates the Mott and metallic phases and the bath hybridizations and energies are independent of $\mu$ within the Mott gap; also, the SFA yields a first-order Mott transition as a function of $U$ in two dimensions, which is not seen in the CDMFT solution of the same bath system.
We benchmarked different weight functions against the solution found in Potthoff's Self-Energy Approach, and argued that the best choices are provided by weight functions that have the lowest SFA gradient.
Weight functions that promote low frequencies in an `exaggerated' way, e.g. as $1/\om$ (\ref{eq:W}b), are less adequate for phases with no or weak spectral gap, because the self-energy is small at low frequencies in those phases.
We proposed a weight function proportional to the self-energy squared (\ref{eq:W}c), inspired by the role played by the self-energy in the SFA variational condition (\ref{eq:EulerVCA}). This new weight function is the most successful at describing the $U$-driven Mott transition in two dimensions.
We also pointed out that the fictitious temperature used in evaluating the distance function should be sufficiently small, $\beta=100/t$ being a good rule-of-thumb value.

\acknowledgements
Fruitful discussions with A.-M.~Tremblay are gratefully acknowledged. Computational resources were provided by RQCHP and Compute Canada.


%


\end{document}